# Low Complexity V-BLAST MIMO-OFDM Detector by Successive Iterations Reduction

Karam Ahmed[1], Sherif Abuelenin[1], Heba Soliman[1], Khairy Al-Barbary[2]

[1]Port-Said University, Faculty of Engineering, 42526, Egypt
[2]Suez Canal University, Faculty of Engineering, 41522, Egypt
karam.ebrahiem@te.eg

*Abstract*—V-BLAST detection method suffers large computational complexity due to its successive detection of symbols. In this paper, we propose a modified V-BLAST algorithm to decrease the computational complexity by reducing the number of detection iterations required in MIMO communication systems. We begin by showing the existence of a maximum number of iterations, beyond which, no significant improvement is obtained. We establish a criterion for the number of maximum effective iterations. We propose a modified algorithm that uses the measured SNR to dynamically set the number of iterations to achieve an acceptable bit-error rate. Then, we replace the feedback algorithm with an approximate linear function to reduce the complexity. Simulations show that significant reduction in computational complexity is achieved compared to the ordinary V-BLAST, while maintaining a good BER performance.

## I. INTRODUCTION

Multiple Input Multiple Output (MIMO) systems can significantly increase the spectral efficiency in wireless communications. The corresponding technology is known as spatial multiplexing, in which, both transmit and receive sides use antenna arrays to transmit different data concurrently [1, 2]. The broadband communication channels are frequency-selective; utilizing Orthogonal Frequency Division Multiplexing (OFDM) overcomes the frequency selective nature of broadband channels, by converting them to multiple flat fading sub-channels [3, 4]. Vertical-Bell Laboratories Layered Space-Time (V-BLAST) architecture is effectively used for detection in MIMO systems. It is based on successive interference cancellation and needs to calculate the pseudo-inverse of the channel matrix and its deflated matrices [5, 6]. Integrating OFDM and MIMO systems rapidly increases the complexity. This is because detection is required for each subcarrier of the OFDM system. V-BLAST requires optimal ordering of the channel matrix to maximize the minimum post-detection signal-to-noise ratio (SNR) of all data streams [7, 8], which means the strongest signal has to be detected first to reduce its effect on the remaining signals [9]. After detecting a sufficient number of the strongest symbols, there is no need to continue in iterations of detection, and the remaining part of signals can be detected directly using a linear detector; e.g. Zero Forcing (ZF) or Minimum Mean Square Error (MMSE) detectors [5]. Other methods have been introduced in order to reduce the complexity of V-BLAST algorithm e.g., [10-12].

In this paper, we introduce a V-BLAST algorithm with reduced complexity. We reach a criterion for the maximum number of effective iterations ($N_{imax}$) in a quadrate MIMO system, (system with equal number of transmitting antennas $N_t$ and receiving antennas $N_r$). Then, we build a complete iterative V-BLAST detector with variable number of successive iterations ($N_i$). We first utilize a feedback based algorithm using the measured SNR value. In each iteration, the algorithm checks the required BER value is reached. If so, the remaining symbols are detected linearly. Otherwise, extra iterations are performed until achieving the target BER, or reaching $N_{imax}$.

To reduce complexity further, we use a linear function to approximate the performance of the dynamic iterative algorithm. The two algorithms are compared with the ordinary V-BLAST detector using two criteria; BER performance and complexity. Complexity is assessed based on the average execution time of the algorithms.

The rest of this paper is organized as follows; section II briefly describes MIMO-OFDM systems, followed by the model of ordinary V-BLAST detection algorithm in section III. Section IV describes the proposed modifications. Finally, the paper is concluded in section V.

## II. MIMO OFDM MODEL

This section presents the model for the $N_r \times N_t$ MIMO system at $N_t = N_r = N_a$. We consider non-line-of-sight (NLOS) Rayleigh flat fading channel.. The entries of the channel matrix (H) are independent and identically distributed (i.i.d) complex Gaussian random variables. The signal is corrupted by i.i.d zero mean additive white Gaussian noise [13, 14]. Block diagram of MIMO-OFDM system for one subcarrier k, where k = 0, 1, …K-1 is shown in Fig. 1 [15].

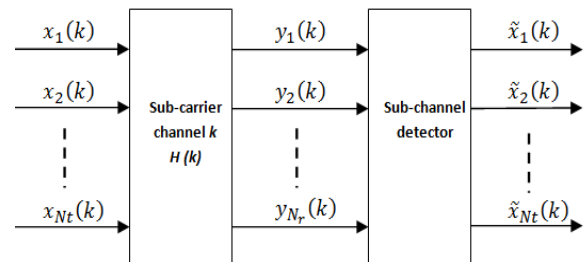

Figure 1. Block diagram of the baseband subcarrier (k) of MIMO-OFDM system.

The received signal is given by:
$$Y(k) = H(k)X(k) + N(k) \qquad (1)$$
Where

$$Y(k) = [y_1(k), y_2(k), \ldots, y_{N_r}(k)]^T \quad (2)$$

And the channel matrix for subcarrier k is:

$$H(k) = \begin{vmatrix} h_{11}(k) & h_{12}(k) & \cdots & h_{1Nt}(k) \\ h_{21}(k) & h_{22}(k) & \cdots & h_{2Nt}(k) \\ \vdots & \vdots & \ddots & \vdots \\ h_{Nr1}(k) & h_{Nr2}(k) & \cdots & h_{NrNt}(k) \end{vmatrix} \quad (3)$$

where, H is a Rayleigh flat fading channel impulse response.

The transmitted (input) signal is:

$$X(k) = [x_1(k), x_2(k), \ldots, x_{N_t}(k)]^T \quad (4)$$

, the added noise is:

$$N(k) = [n_1(k), n_2(k), \ldots, n_{N_r}(k)]^T \quad (5)$$

and the estimated signal is:

$$\widetilde{X}(k) = [\widetilde{x}_1(k), \widetilde{x}_2(k), \ldots, \widetilde{x}_{N_t}(k)]^T \quad (6)$$

## III. NON-LINEAR V-BLAST DETECTOR

V-BLAST nonlinear detection algorithm [15] provides a good BER performance compared to ZF and MMSE. It is regarded as an Ordered Successive Interference Cancellation (OSIC) detection scheme, where the output signals are detected successively combined with spatial nulling process [16]. In an ($N_r$x$N_t$) MIMO system, we need $N_t$ detection steps, corresponding to the number of transmit antennas. The main problem with V-BLAST is the error propagation. To circumvent error propagation, detection is achieved in a descending order starting from the strongest sub-stream. The nulling process plays the main role in detection. In each detection iteration, the detector treats the other sub-streams as interference signals. After detecting the strongest sub-stream, its contribution is subtracted from the received signal. The remaining part contains the weak signals that can be recovered in the next iteration without the interference effect of the strongest one. This procedure is repeated until detecting all sub-streams. Both ZF, MMSE can be used to provide the spatial nulling process and to separate the individual streams. Finally, a quantization process is applied to approximate each estimated signal to the nearest value in the constellation [1].

The general algorithm for V-BLAST detection is as follows [17, 18].

*Ordering*: by choosing the row $w_{ki}$ of $w(k)$ with the minimum Euclidean norm.

$$w_{ki} = (G_i)_k \quad (7)$$

*Nulling*: where we use the null vector $w_{ki}$ to null out the undesired signals and obtain the required one.

$$Z_{ki} = w_{ki} Y_i \quad (8)$$

*Slicing*: approximate the detected signal to its nearest value in signal constellation.

$$\hat{X}_{bi} = O(Z_{ki}) \quad (9)$$

*Cancellation*: Subtract the detected signal from the received signal vector and zeroing its corresponding column in channel matrix H.

$$Y_{i+1} = Y_i - (H_i)^{k_i} \hat{X}_{k_i} \quad (10)$$

$$H_{i+1} = H_i - (H_i)^{K_i} \quad (11)$$

Finally the detected symbols are exported.
*Export:*

$$\hat{X} = [\hat{X}_1, \hat{X}_2, \ldots, \hat{X}_M]^T \quad (12)$$

where G = Pinv(H) for ZF-V-BLAST, and reordering is done with the smallest row norm of G. If we use MMSE linear detector, we have $G = DH^H$ [19], where

$$D = [H^H H + I_{Nt} \frac{1}{SNR}]^{-1} \quad (13)$$

And we do the reordering at the smallest diagonal entry of D [20]. $(G_i)_j, (H_i)_{bi}$ indicate the jth row of $G_i$ and the bi column of $H_i$, respectively.

## IV. METHODOLOGY

### A. Estimating Nimax

V-BLAST detects stronger symbols first. Detecting the remaining weak signals successively does not provide a significant performance improvement. To show this, we vary the number of iterations from zero (linear detector) to ($N_t$-1) iterations (ordinary V-BLAST). The remaining signals in each time are detected using a linear method. We perform this routine on different MIMO antenna configurations at different coding schemes.

Figs. (2 through 4) show the results for 4x4, 8x8 and 16x16 MIMO systems using QPSK and 16QAM modulation schemes. Both ZF and MMSE linear detectors are used as the core of our V-BLAST detector.

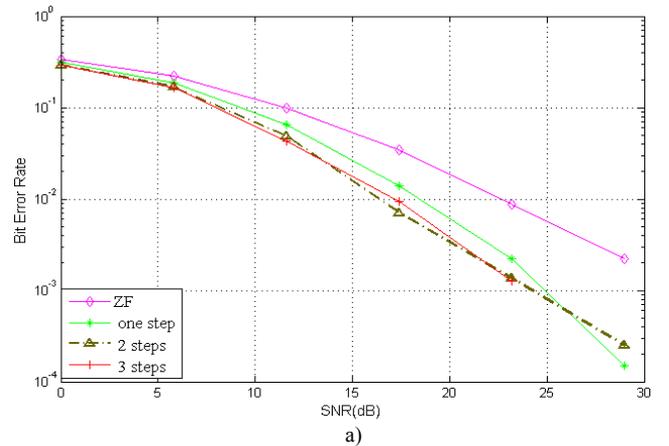

a)

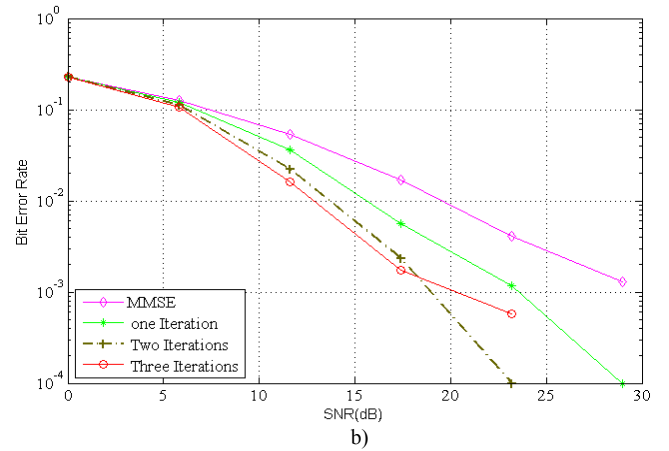

b)

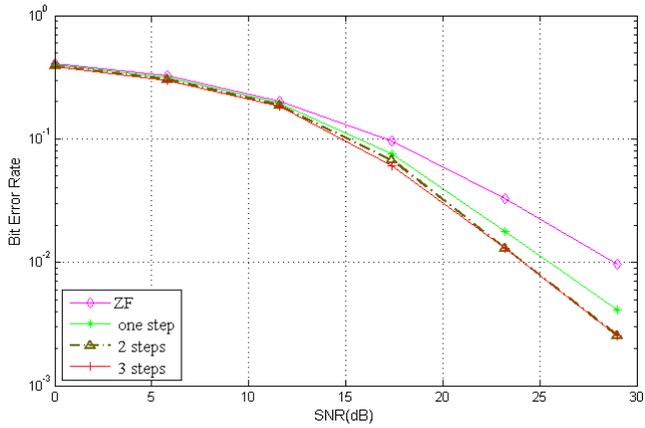

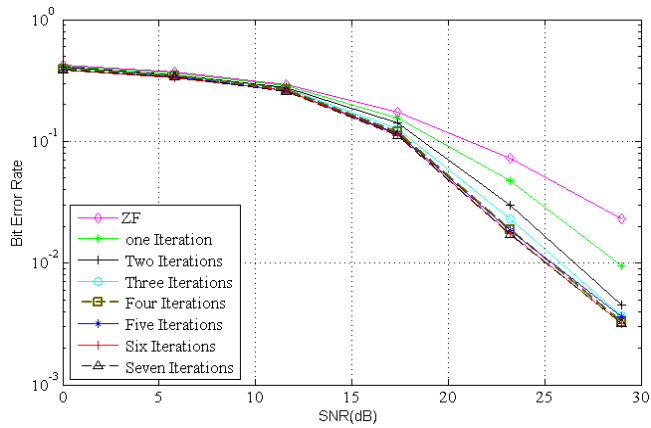

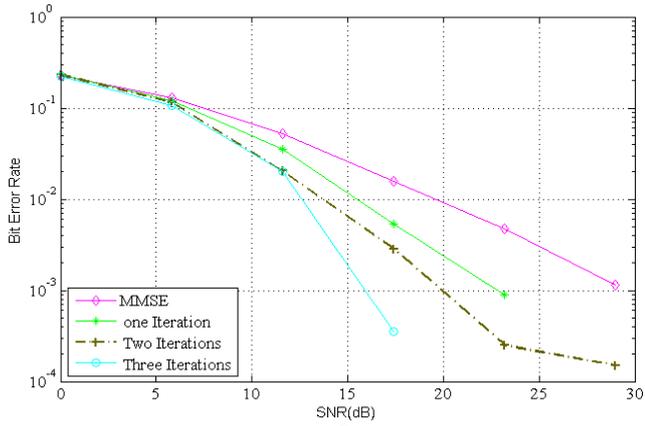

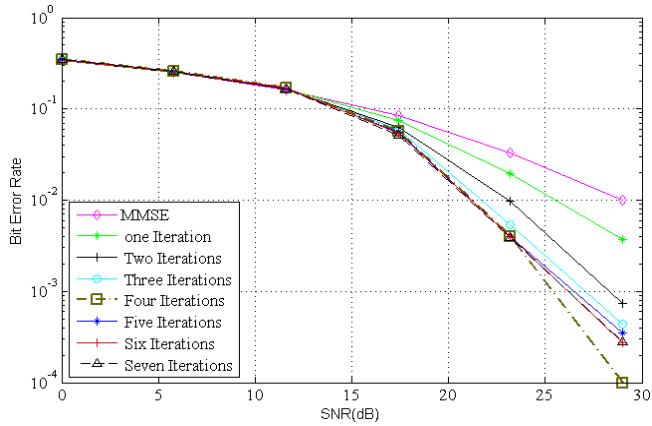

Figure 2. The BER performance for 4x4 MIMO system using the variable iterations V-BLAST detector a) ZF/QPSK, b) MMSE/QPSK, c) ZF/16QAM, and d) MMSE/16QAM

Figure 3. The BER performance for 8x8 MIMO system using the variable iterations V-BLAST detector a) ZF/QPSK, b) MMSE/QPSK, c) ZF/16QAM, and d) MMSE/16QAM

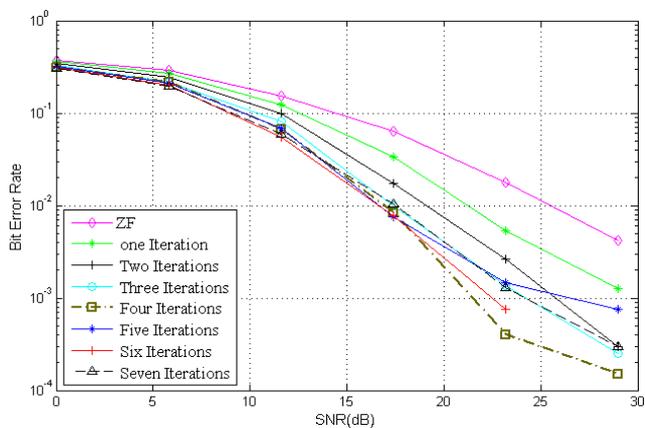

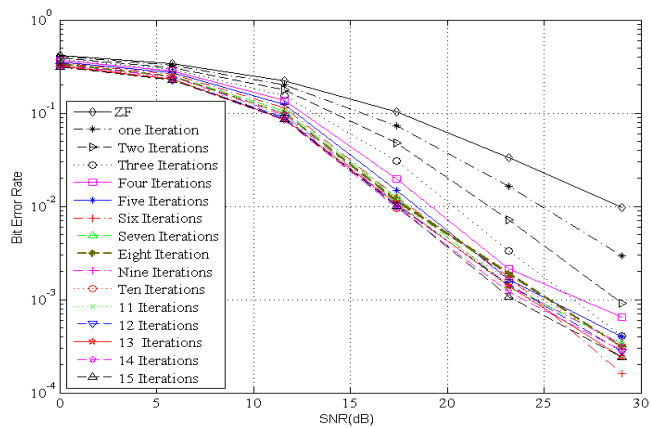

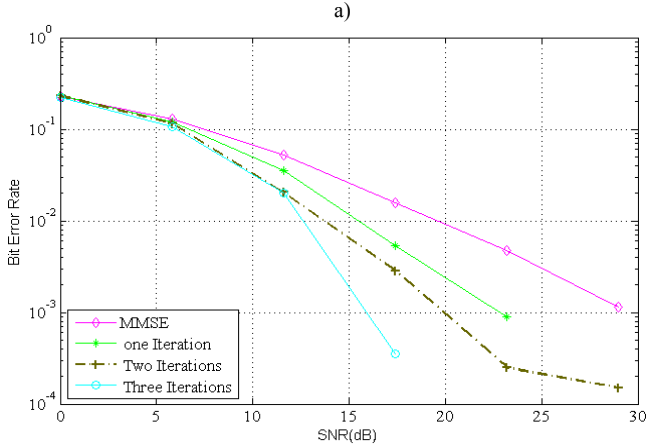

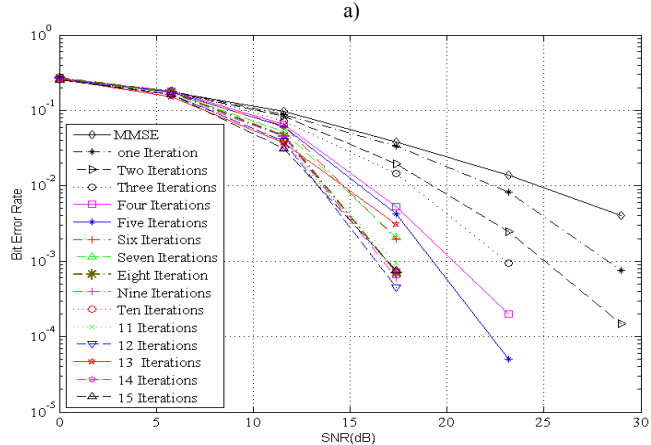

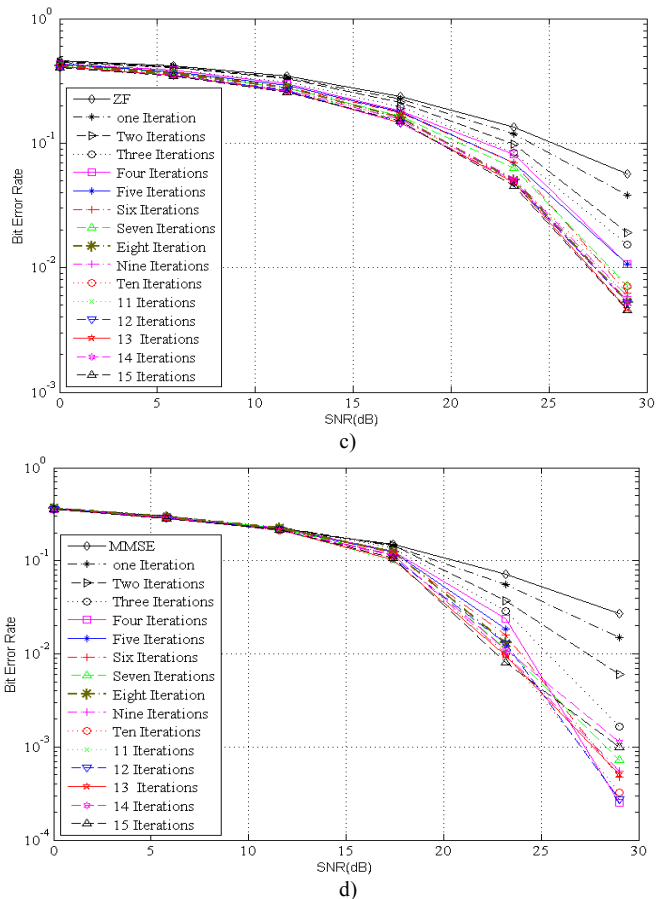

Figure 4. The BER performance for 16x16 MIMO system using the variable iterations V-BLAST detector a) ZF/QPSK, b) MMSE/QPSK, c) ZF/16QAM, and d) MMSE/16QAM

We can clearly see that for $Ni > Nt/2$, no significant improvement in performance is obtained for all studied configurations. MMSE is preferred as a linear detector for its superior performance over ZF [21, 22].

### B. V-BLAST with variable iterations (Ni)

For further complexity reduction, we dynamically set the number of iterations based on the measured (estimated) SNR value. Different methods may be utilized for SNR estimation, e.g. the pilot preamble scheme (data-aided) and semi-blind SNR Estimation (none-data-aided). The blind scheme can be applied simply on MIMO-OFDM systems, where the scheme can be used directly before or after the MIMO equalizer [23].

Our algorithm is applied on a 16QAM-8X8 MIMO system. The initial modification is shown in Fig. 5 a). Starting from the first iteration, the algorithm checks if the estimated SNR leads to the accepted BER. If so, the remaining symbols are detected linearly. Otherwise, the iterations continue until reaching the BER requirement or reaching $Nimax$. The drawback of this method is that it actually increases the complexity for low values of SNR. This is mainly because the nulling, slicing and cancellation steps are repeated Ni+1 times.

The second algorithm, with flowchart shown in Fig. 5 b), uses a formula, to directly determine the number of iterations, $Ni$, from the SNR value.

Assuming a maximum required BER of 1E-2 in an 8x8 16-QAM MIMO system, with SNR ranging between 16 dB and 34dB [24]. First, we simulate the V-BLAST starting from $Ni=0$ to $Ni=Nimax$, as shown in Fig. 6 a). Table I lists the required $Ni$ values for the defined acceptable performance under different SNR conditions. For simplicity, we approximate the relation between SNR and $Ni$ as a linear formula, as shown in Fig. 6 b).

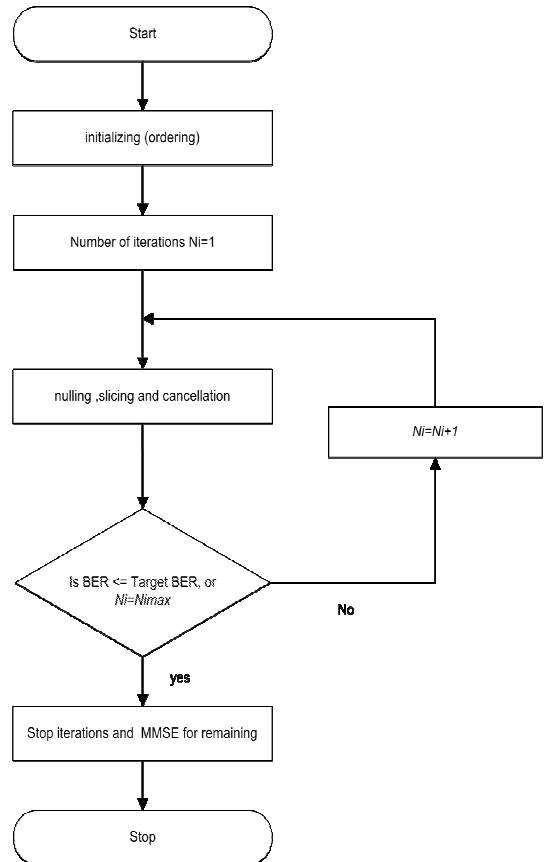

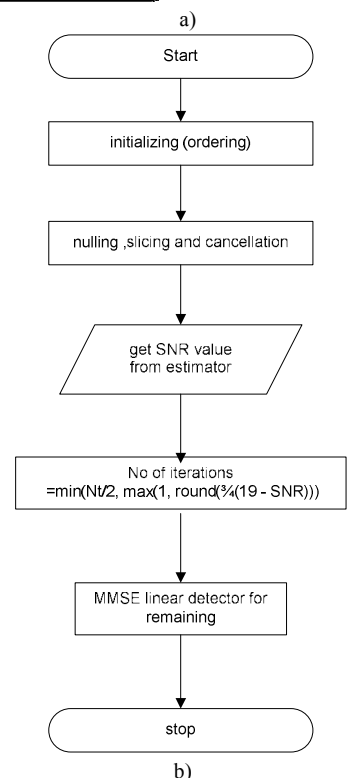

Figure 5. The flow chart of the proposed method for 8x8 MIMO systems based on the feedback and the formula methods a) The feedback method and b) The formula method

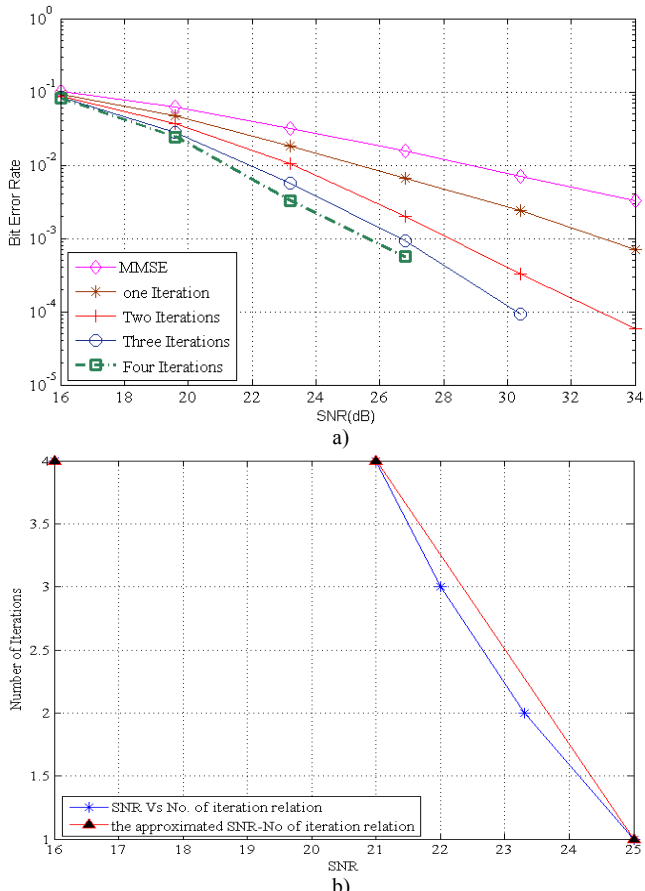

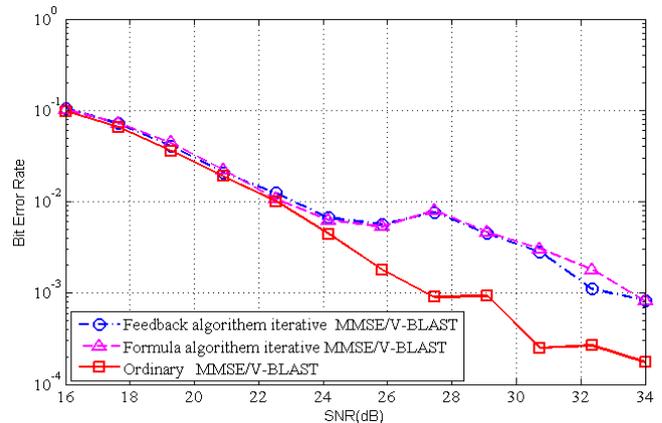

Figure 6. a) The plot of Bit Error Rate of 16QAM MMSE/V-BLAST at Ni from 0 to Nt-1 b) The plot of the relation between the number of V-BLAST iterations and the SNR in dB under the condition of (BER <= 1E-2)

TABLE I. NUMBER OF ITERATION AND ITS CORRESPONDING SNR VALUES FOR PROPOSED CONDITION

| Number of iterations | 4 | 4 | 3 | 2 | 1 | 1 |
|---|---|---|---|---|---|---|
| SNR value in dB | 16 | 21 | 22 | 23.3 | 25 | 34 |

The relation between Ni and SNR can be approximated as (the linear curve in Fig 6. b));

$$\therefore Ni = \min(\frac{N_t}{2}, \max(1, round(\frac{3}{4}(19-SNR)))) \quad (14)$$

The *round* function rounds the obtained value to the nearest integer number. Also to maintain the iterations between 1 and $N_t/2$, we use minimum and maximum operators.

Note that, we can change the value of the minimum accepted SNR as required to comply with different practical situations.

Table II shows that the formula method has a significant reduction in complexity over the feedback method and over the ordinary V-BLAST. Ordinary V-BLAST significantly outperformed the modified algorithm for SNR values greater than 24 dB, but with the price of unnecessary added complexity after reaching the required BER performance. While, Fig. 7 shows that the two proposed algorithms almost have the same performance.

TABLE II. AVERAGE COMPLEXITY RATIO FOR EACH V-BLAST DETECTION TYPE WITH RESPECT TO ORDINARY V-BLAST OF EACH DETECTION ALGORITHEM

| Detection Algorithm | Ratio of Complexity |
|---|---|
| The Ordinary V-BLAST | 100% |
| The feedback algorithm | 122% |
| Fixed Number of Iterations at $Ni=Nimax$ | 74.2% |
| The formula algorithm | 57% |

Figure 7. Comparison of BER performance of the two proposed algorithms with each other and the ordinary V-BLAST

## V. CONCLUSION

In this paper, we established a criterion of the maximum efficient number of iterations in V-BLAST; we showed that for $N > N_t/2$ (at $N_r=N_t$) no significant improvement is obtained. For complexity reduction, we proposed a variable iteration V-BLAST algorithm within the maximum limit of iterations. We obtained a complexity reduction of about 43% compared to the ordinary V-BLAST algorithm without a significant degradation in BER in the SNR range of interest. A practical implementation is considered in future work.